\documentclass[preprintnumbers,showpacs,amsmath,amssymb,floatfix,10pt,prd,onecolumn,
nofootinbib]{revtex4}
\usepackage{latexsym}
\usepackage{epsfig}
\usepackage{amssymb}

\begin{document}

\title{Bianchi-V string cosmology with power law expansion in $f(R, T)$ Gravity}

\author{Anil Kumar Yadav}
\email{abanilyadav@yahoo.co.in} \affiliation{Department of Physics,
Anand Engineering College, Keetham, Agra 282 007, India}

\begin{abstract}
{\bf Abstract:}  In this paper, we search the existence of Bianchi-V string cosmological model in $f(R,T)$ 
gravity with power law expansion. Einstein's field equations have been solved by taking into account the 
law of variation of Hubble's parameter that yields the constant value of deceleration parameter (DP). 
We observe that the massive strings dominate the early universe but they do not survive for long time 
and finally disappear from the universe. We examine the nature of classical potential and also discuss the 
physical properties of universe.\\


\end{abstract}


\maketitle

\section{Introduction}
The spontaneous breakdown of symmetries in the early universe that produce linear 
discontinuities in the field is called cosmic string (Kibble 1976). The cosmic strings are 
also common in modern string cosmologies. In 2003, general interest in cosmic strings 
was heightened by the discovery of what seemed at first to be a plausible candidate 
for lensing by a cosmic string. A pair of images of elliptical galaxies separated by 1.8 
arc seconds was found to have the same redshift, $z = 0.46$, and the same spectra. Sazhim et al (2003) 
confirmed that these images were not distorted in the way that would be expected for lensing of a single 
galaxy but are consistent with lensing by cosmic string. In the literature,
the line like structure of cosmic strings with particle attached to them are considered as possible seeds 
for galaxy formation at the early stages of the evolution of universe. In the past time, Stachel (1980), Letelier (1983) 
and Vilenkin et al (1987) have studied different aspects of string cosmological models in general relativity. 
Reddy (2003), Reddy and Naidu (2007) and recently Yadav (2013) have investigated anisotropic string cosmological 
models in scalar-tensor theory of gravitation.\\ 

Harko et al (2011) proposed $f(R,T)$ gravity theory by taking into account the gravitational 
Lagrangian as the function of Ricci scalar R and of the trace of energy-stress tensor T. They have obtained 
the equation of motion of test particle and the gravitational field equation in metric formalism both. 
Myrzakulov (2011) presented point like Lagrangian's for $f(R,T)$ gravity. The $f(R,T)$ gravity model that satisfy 
the local tests and transition of matter from dominated era to accelerated phase was considered by Houndjo (2012). 
Recently Chaubey and Shukla (2013), Naidu et al (2013) and Ahmad and Pradhan (2013) have investigated anisotropic 
cosmological model in $f(R,T)$ gravity. In this paper, our aim is to study the role of strings in $f(R,T)$ gravity 
and Bianchi-V space time. Since Bianchi-V models are natural generalization of FRW models hence the Bianchi-V 
cosmological models permit one to obtain more general cosmological model, in comparison to FRW model. In the 
recent years several authors (Yadav 2011, Kumar and Yadav 2011, Yadav 2012, Pradhan et al 2005, Singh et al 2008) 
have studied Bianchi-V cosmological models in different physical context.\\ 

In this paper, we establish the existence of Bianchi-V string cosmological model in $f(R,T)$ gravity and examine 
a cosmological scenario by proposing power law expansion. We observe that strings do not survive for long time and 
eventually disappear from universe. The organization of the paper is as follows: The model and basic theory of $f(R,T)$ gravity are presented in section 2. 
The field equations are established in section 3 and their solution is presented in section 4. Section 5 deals with 
the cosmological parameters and classical potential of derived model. Finally conclusions are presented in section 6.\\

\section{The metric and $f(R,T)$ gravity}
We consider the Bianchi-V metric in following form
\begin{equation}
 \label{eq1}
ds^2 = -dt^2+A^{2}dx^{2}+e^{2\alpha x}(B^{2}dy^{2}+C^{2}dz^{2})
\end{equation}
Here, $A(t)$, $B(t)$ \& $C(t)$ are scale factors in the x, y \& z direction respectively and $\alpha$ is constant.\\

The action of $f(R,T)$ gravity is given by
\begin{equation}
 \label{eq2}
S=\frac{1}{16\pi}\int{f(R,T)\sqrt{-g}d^{4}x}+\int{L_{m}\sqrt{-g}d^{4}x}
\end{equation}
Where $R$, $T$ and $L_{m}$ are the Ricci scalar, the trace of the stress-energy tensor of matter and the 
matter Lagrangian density respectively.\\
The stress-energy tensor of matter is given by
\begin{equation}
 \label{eq3}
T_{ij}=-\frac{2}{\sqrt{-g}}\frac{\delta \sqrt{-g}L_{m}}{\delta g^{ij}}
\end{equation}
The gravitational field equation of $f(R,T)$ gravity is given by
\begin{equation}
 \label{eq4}
f_{R}(R,T)R_{ij}-\frac{1}{2}f(R,T)g_{ij}+(g_{ij}\triangledown^{i}\triangledown_{i}-
\triangledown_{i}\triangledown_{j})f_{R}(R,T)= 8\pi T_{ij}-f_{T}(R,T)T_{ij}-f_{T}(R,T)\Theta_{ij}
\end{equation}
where $f_{R}(R,T) = \frac{\partial f(R,T)}{\partial R}$, $f_{T} = \frac{\partial f(R,T)}{\partial T}$, 
$\Theta_{ij} = -2T_{ij} - pg_{ij}$ and $\triangledown_{i}$ denotes the covariant derivative.\\
The stress-energy tensor is given by 
\begin{equation}
 \label{eq5}
T_{ij}=(\rho+p)u_{i}u_{j}-pg_{ij}
\end{equation}
In general, the field equations depend through the tensor $\Theta_{ij}$, on the physical 
nature of the matter field. Hence we obtain several theoretical models for different choice of 
f(R,T) depending on the nature of the matter source. In the literature, Chaubey $\&$ Shukla (2013), 
Reddy et al. (2012a, 2012b) and Naidu et al (2013) have been studied the cosmological models, assuming $f(R,T) = R + 2f(T)$. 
Recently Ahmad $\&$ Pradhan studied consequence of Bianchi-V cosmological model in f(R,T) gravity by considering 
$f(R,T) = f_{1}(R)+f_{2}(T)$. They have assumed perfect fluid as source of matter while in this paper, we assumed 
the string fluid as source of matter to describe the physical consequences of early universe. Thus our paper is 
all together different from  the paper of Ahmad and Pradhan (2013). Assuming 
$f_{1}(R) = \mu R$ and $f_{2}(T) = \mu T$ where $\mu$ is arbitrary parameter.\\
Now equation (\ref{eq4}) can be rewritten as
\begin{equation}
\label{eq6}
R_{ij}-\frac{1}{2}g_{ij}R = \left(\frac{8\pi +\mu}{\mu}\right)T_{ij}+\left(p+\frac{1}{2}T\right)g_{ij}
\end{equation}
Throughout the paper, we use units $c = G = 1$.\\
The expansion scalar $(\theta)$ and shear scalar $(\sigma)$ have the form
\begin{equation}
 \label{eq7}
\theta = 3H = \frac{\dot A}{A} + \frac{\dot B}{B} + \frac{\dot C}{C}
\end{equation}
\begin{equation}
 \label{eq8}
2\sigma^{2} = \left[\frac{\dot A^{2}}{A^{2}}+\frac{\dot B^{2}}{B^{2}}+\frac{\dot C^{2}}{C^{2}}\right] - 
\frac{\theta^{2}}{3}
\end{equation}

\section{Field equations  in $F(R,T)$ gravity}
The Einstein's field equation (\ref{eq6}) for the line-element (\ref{eq1}) 
leads to the following system of equations
\begin{equation}
 \label{eq9}
\frac{\ddot{B}}{B}+\frac{\ddot{C}}{C}+\frac{\dot{B}\dot{C}}{BC}-\frac{\alpha^{2}}{A^{2}} = -
\left(\frac{8\pi+\mu}{\mu}\right)\left(p + \lambda\right)
\end{equation}
\begin{equation}
 \label{eq10}
\frac{\ddot{A}}{A}+\frac{\ddot{C}}{C}+\frac{\dot{A}\dot{C}}{AC}-\frac{\alpha^{2}}{A^{2}} = 
-\left(\frac{8\pi+\mu}{\mu}\right)p 
\end{equation}
\begin{equation}
 \label{eq11}
\frac{\ddot{A}}{A}+\frac{\ddot{B}}{B}+\frac{\dot{A}\dot{B}}{AB}-\frac{\alpha^{2}}{A^{2}} = 
-\left(\frac{8\pi+\mu}{\mu}\right)p 
\end{equation}
\begin{equation}
 \label{eq12}
\frac{\dot{A}\dot{B}}{AB}+\frac{\dot{A}\dot{C}}{AC}+\frac{\dot{B}\dot{C}}{BC}-\frac{3\alpha^{2}}{A^{2}} = 
\left(\frac{8\pi+\mu}{\mu}\right)\rho
\end{equation}
\begin{equation}
 \label{eq13}
\frac{2\dot{A}}{A}-\frac{\dot{B}}{B}-\frac{\dot{C}}{C} = 0
\end{equation}

\section{solution of field equation}
We have system of five equation (\ref{eq9})$-$(\ref{eq13}) involving six unknown variables, 
namely $A$, $B$, $C$, $\rho$, $p$ \& $\lambda$. Thus, in order to completely solve the fields 
equations, we need at least one physical assumption among the unknown parameters. In the literature,
it is common to use the law of variation of Hubble's parameter which yields the constant value of 
deceleration parameter. This law deals with two type of cosmology (i) power law cosmology (ii) 
exponential law cosmology. It is well known that the exponential law projects the dynamics of 
future universe and such type of model does not have consistency with present day observations. 
Since we are looking for a model, describing the late time acceleration of universe therefore we 
choose the average scale factor in following form
\begin{equation}
 \label{eq14}
a = (nDt)^{1/n}
\end{equation}
where $n \neq 0$, is positive constant.\\
Eq. (\ref{eq14}) can be easily obtain by law of variation of Hubble's 
parameter.\\ 
We define average scale factor $(a)$ and mean Hubble's parameter $(H)$ of the 
Bianchi-V model as
\begin{equation}
 \label{eq15}
a = (ABC)^{1/3}
\end{equation}
\begin{equation}
 \label{eq16}
H = \frac{\dot{a}}{a} = \frac{1}{3}(H_{1}+H_{2}+H_{3})
\end{equation}
where $H_{1} = \frac{\dot{A}}{A}$, $H_{2} = \frac{\dot{B}}{B}$ and $H_{3} = \frac{\dot{C}}{C}$ 
are directional Hubble's parameters in the direction of x, y, and z- direction respectively.\\
From equation (\ref{eq10}) and (\ref{eq11}), we obtain the following relation
\begin{equation}
 \label{eq17}
\frac{B}{C} = b_{1}exp\left(x_{1}\int\frac{dt}{ABC}\right)
\end{equation}
where $b_{1}$ and $x_{1}$ are constant of integrations.\\
Integrating eq. (\ref{eq13}), we get
\begin{equation}
 \label{eq18}
A^{2} = BC
\end{equation}
From equations (\ref{eq14}), (\ref{eq15}), (\ref{eq17}) and (\ref{eq18}), the metric functions can 
be explicitly written as
\begin{equation}
 \label{eq19}
A = (nDt)^{\frac{1}{n}}
\end{equation}
\begin{equation}
 \label{eq20}
B = \sqrt{b_{1}}(nDt)^{\frac{1}{n}}exp\left[\frac{x_{1}}{2D(n-3)}(nDt)^{\frac{n-3}{3}}\right]
\end{equation}
\begin{equation}
 \label{eq21}
C = \frac{1}{\sqrt{b_{1}}}(nDt)^{\frac{1}{n}}exp\left[-\frac{x_{1}}{2D(n-3)}(nDt)^{\frac{n-3}{3}}\right]
\end{equation}
provided $n \neq 3$.
\section{The cosmological parameters and classical potential}
The average Hubble's parameter $(H)$, expansion scalar $(\theta)$, spatial volume $(V)$,  
deceleration parameter (DP) and shear scalar $(\sigma)$ are given by
\begin{equation}
 \label{eq22}
H = \frac{1}{nt}
\end{equation}
\begin{equation}
 \label{eq23}
\theta = 3H = \frac{3}{nt}
\end{equation}
\begin{equation}
 \label{eq24}
V = (nDt)^{\frac{3}{n}}
\end{equation}
\begin{equation}
 \label{eq24}
q = -\frac{a\ddot{a}}{{\dot{a}}^2} = n-1
\end{equation}
\begin{equation}
 \label{shear}
\sigma = \frac{nx_{1}}{6}(nDt)^{\frac{n-6}{3}}
\end{equation}

For accelerating universe, we impose the restriction on the value of n $(0 < n < 1)$. If we take $n = 0.27$ then 
the value of DP is $-0.73$ which exactly matches with the observed value of DP at present epoch (Cunha et al. 2009). 
Hence we constrain $n = 0.27$ in graphical representations and discussion of physical parameters of derived model.\\
The isotropic pressure $(p)$, proper energy density $(\rho)$, string tension density $(\lambda)$ 
and particle energy density $(\rho_{p})$ are found to be
\begin{equation}
 \label{eq25}
p = \frac{\mu}{(8\pi +\mu)}\left[\frac{\alpha^2}{(nDt)^{\frac{2}{n}}}+\frac{2n-3}{n^{2}t^{2}}-\frac{7x_{1}}{6t}
(nDt)^{\frac{n-6}{3}}-\frac{n^{2}x_{1}^{2}}{36}(nDt)^{\frac{2(n-6)}{3}}-\frac{n^{2}(n-6)Dx_{1}}{18}(nDt)^{\frac{n-9}{3}}
\right]
\end{equation}
\begin{equation}
 \label{eq26}
\rho = \frac{\mu}{(8\pi +\mu)}\left[\frac{3}{n^{2}t^{2}}-\frac{n^{2}x_{1}^{2}}{36}(nDt)^{\frac{2(n-6)}{3}}
-3\alpha^{2}(nDt)^{-\frac{2}{n}}\right]
\end{equation}
\begin{equation}
 \label{eq27}
\lambda = \frac{\mu}{(8\pi +\mu)}\left[\frac{7x_{1}}{6t}(nDt)^{\frac{n-6}{3}}+\frac{n^{2}(n-6)Dx_{1}}{18}
(nDt)^{\frac{n-9}{3}}-\frac{2(2n-3)}{n^{2}t^{2}}\right]
\end{equation}
\begin{equation}
 \label{eq28}
\rho_{p} = \frac{\mu}{(8\pi +\mu)}\left[\frac{4n-3}{n^{2}t^{2}}-\frac{n^{2}x_{1}^{2}}{36}(nDt)^{\frac{2(n-6)}{3}}
-\frac{7x_{1}}{6t}(nDt)^{\frac{n-6}{3}}-\frac{n^{2}(n-6)Dx_{1}}{18}(nDt)^{\frac{n-9}{3}}-3\alpha^{2}(nDt)^{-\frac{2}{n}}
\right]
\end{equation}

\begin{figure*}[thbp]
\begin{tabular}{rl}
\includegraphics[width=7.5cm]{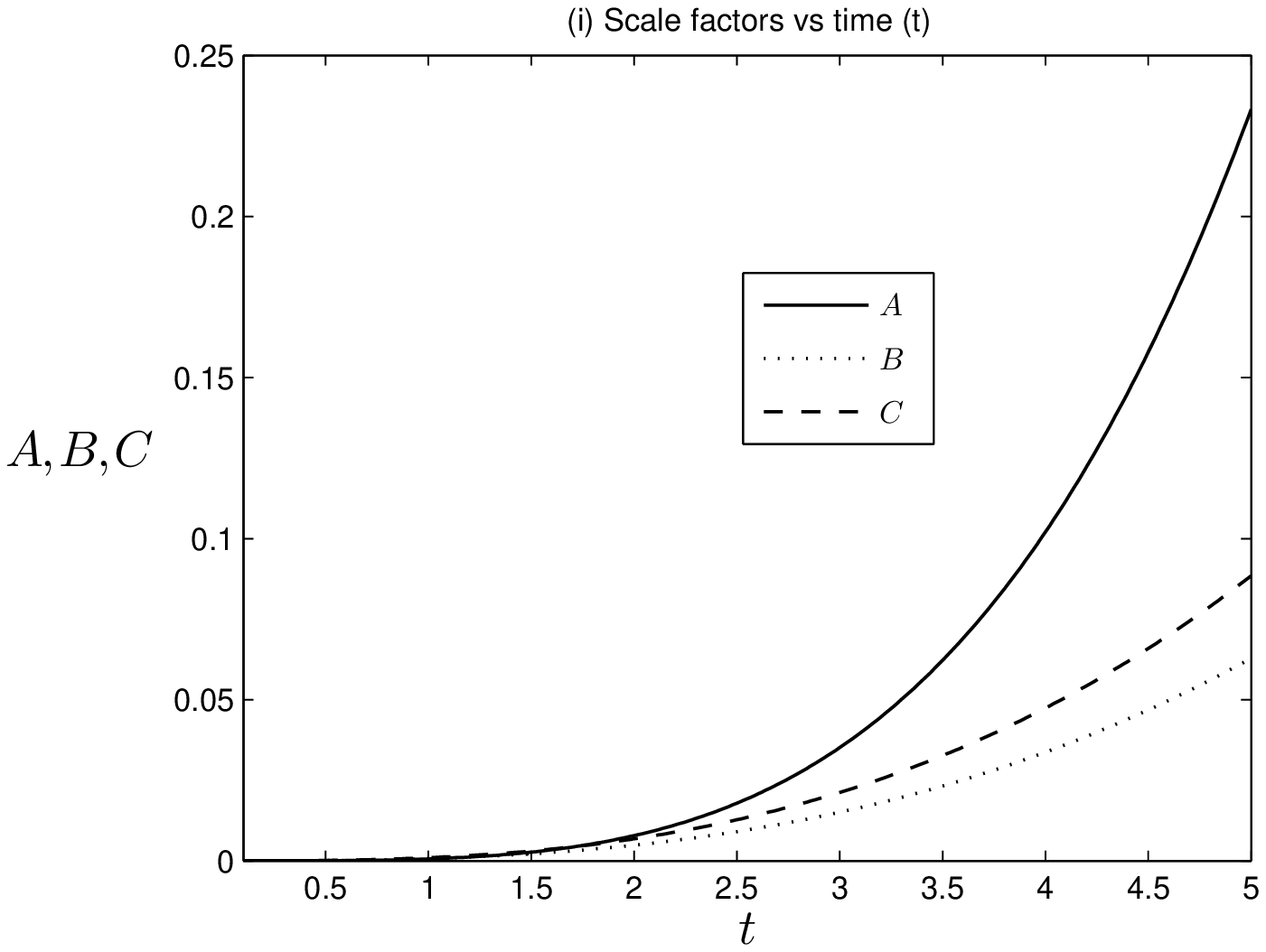}&
\includegraphics[width=7.5cm]{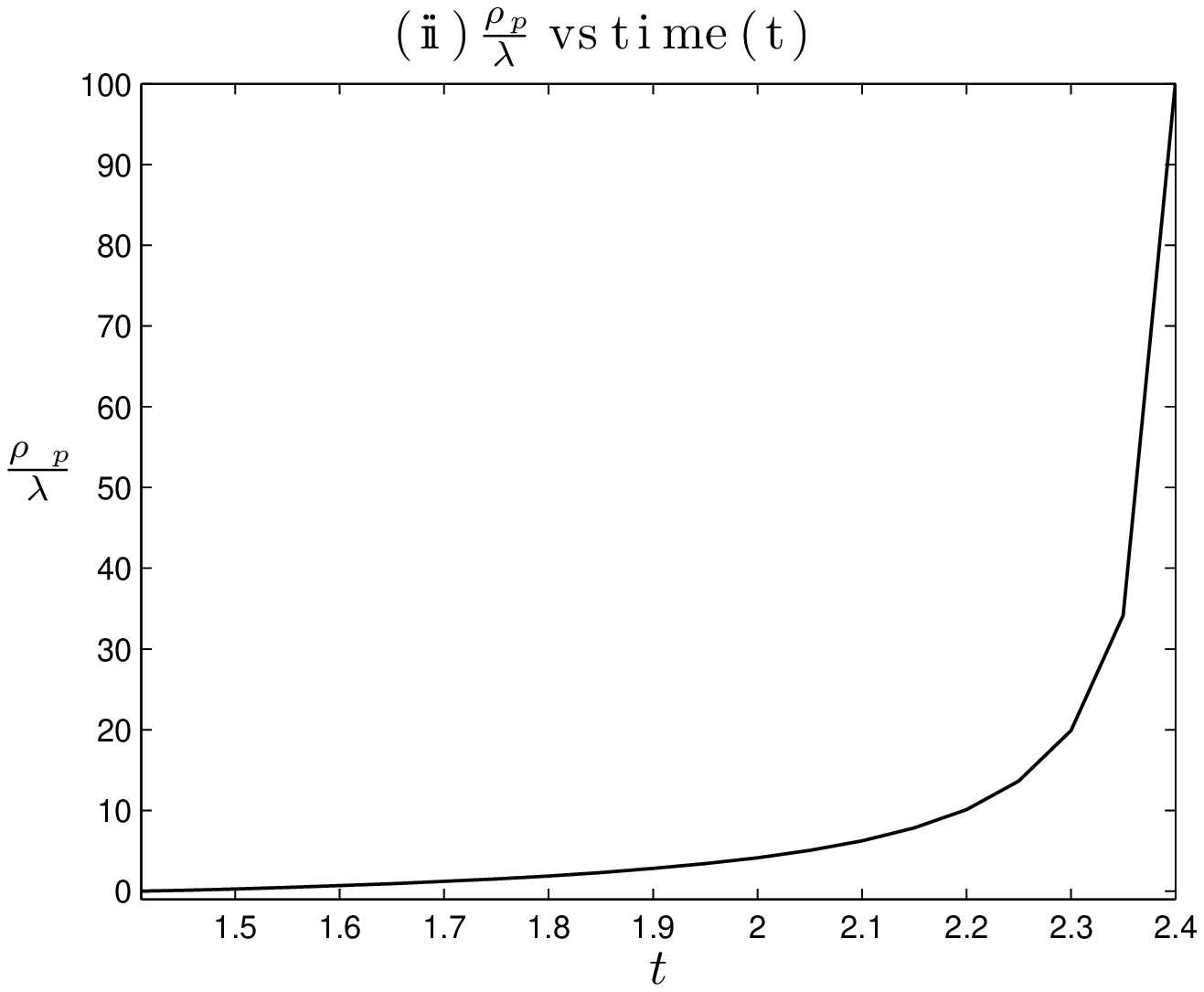} \\
\includegraphics[width=7.5cm]{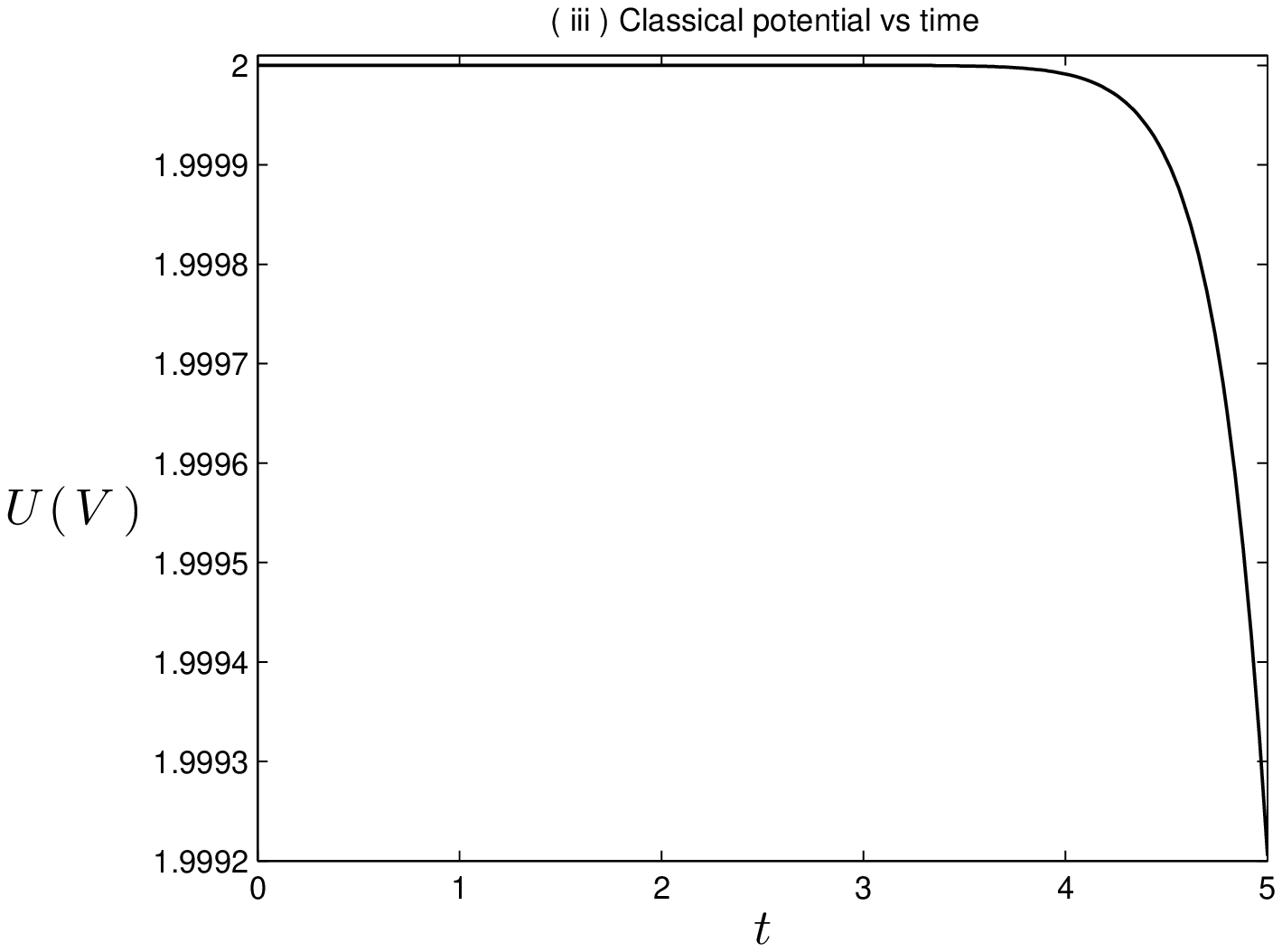}&
\includegraphics[width=7.5cm]{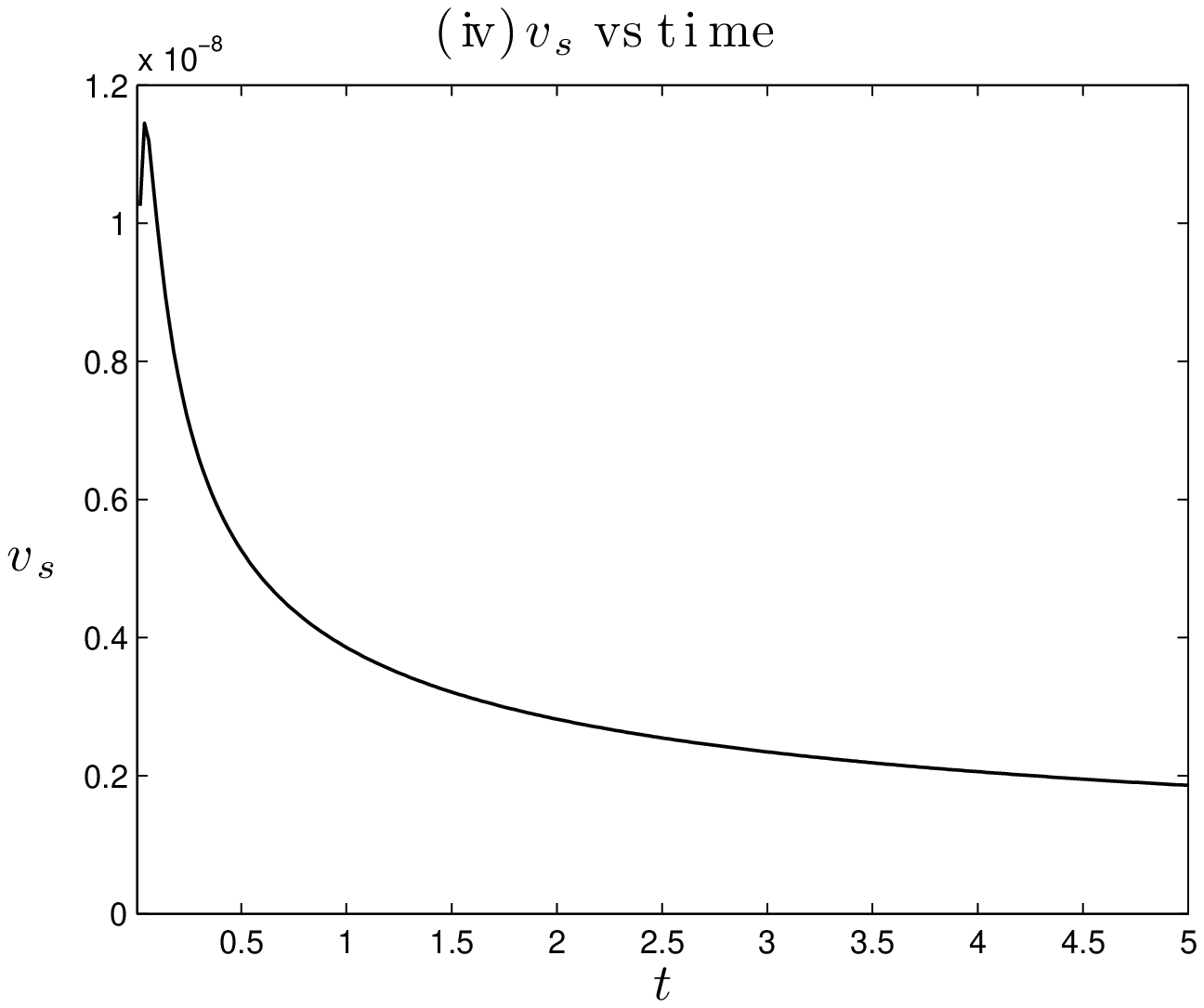}
\end{tabular}
\caption{ Graphs for the case $n=0.27$. }
\end{figure*}
From eq. (\ref{eq17}), we obtain
\begin{equation}
 \label{eq29}
\dot{V} = 3D(nDt)^{\frac{3}{n}-1}
\end{equation}
Following Saha and Boyadjiev (2004) and Yadav(2013), we have the following equation of motion 
of a single particle with unit mass under force $F(V)$ 
\begin{equation}
 \label{eq30}
\dot{V}=\sqrt{2(\epsilon-U(V))}
\end{equation}
Where $U(V)$ and $\epsilon$ are the classical potential and amount of energy respectively.\\
Eqs. (\ref{eq29}) and (\ref{eq30}) lead to
\begin{equation}
 \label{eq31}
U(V) = 2\epsilon-9D^{2}(nDt)^{\frac{2(3-n)}{n}}
\end{equation}
The classical potential in terms of Hubble's parameter is given by
\begin{equation}
 \label{eq32}
U(V)=2\epsilon-9D^{\frac{6}{n}}H^{\frac{2n-6}{n}}
\end{equation}
The age of universe in connection with DP is given by
\begin{equation}
 \label{eq33}
T=\frac{1}{q+1}H^{-1}
\end{equation}
From eq. (\ref{eq33}), it is clear that the value of $q$ in the range $-1 < q < 0$ increase the age of universe.\\
Also we know that the speed of sound $v_{s}$ is less than the speed of light $(c)$. In gravitational unit 
we take $c = 1$. Therefore for a physical viable model, $v_{s}$ lies between 0 and 1.\\
From eq. (\ref{eq25}) and (\ref{eq26}), the speed of sound is given by

\begin{equation}
 \label{eq34}
 v_{s}^{2} = \frac{dp}{d\rho}= \frac{2D\alpha^2(nDt)^{\frac{-(n+2)}{n}}+\frac{4n-6}{n^{2}t^{2}}+\zeta_{1}t^{\frac{n-12}{3}}
+\zeta_{2}(nDt)^{\frac{2n-15}{3}}+
\zeta_{3}(nDt)^{\frac{n-12}{3}}}{\frac{6}{n^{2}t^{2}}+\frac{n^{3}Dx1^{3}}{36}(nDt)^{\frac{2n-15}{3}}-
6D\alpha^{2}(nDt)^{\frac{-(n+2)}{n}}}
\end{equation}
where\\
$\zeta_{1}=\frac{7n(n-9)Dx1}{18}(nD)^{\frac{n-6}{3}}$\\
$\zeta_{2}=\frac{n^{3}(2n-12)dx_{1}^{2}}{108}$\\
$\zeta_{3}=\frac{n^{3}(n-6)(n-9)D}{54}$\\
In figure panel 1, we graphed the parameters of derived model against $t$ for $n = 0.27$. 
The behaviour is quite evident; the scale factors along axial direction satisfies the anisotropic 
nature of universe; $\frac{\rho_{p}}{\lambda  } > 1$ shows that the particles dominate over the 
strings with the evolution of universe hence the strings are not observed today; the classical 
potential $(U(V))$ is positive and speed of sound $(v_{s})$ is less the speed of light throughout the 
expansion of universe from big bang to present epoch.\\     
\section{Concluding remarks}
In the present paper, we have considered $f(R,T)$ gravity model with an arbitrary coupling between 
matter and geometry in Bianchi-V space-time. We have derived the gravitational field equations 
for string fluid corresponding to $f(R,T)$ gravity model. We observed that string tension density $(\lambda)$ 
decreases with time and it approaches to zero at present epoch. Therefore strings could not survive with the 
evolution process of universe. That is why the strings are not observed today but it play significant role 
in early universe. The classical potential $(U(V))$ is positive and it decreases with time. The scale 
factors vanish at $t = 0$. Thus the model has point type singularity at $t = 0$. As $t \rightarrow \infty$ 
the scale factors diverge and the physical parameters such as expansion, 
scalar $(\theta)$, energy density $(\rho)$ and Hubble's parameters $(H)$ tend to zero. 
Therefore in the derived model, all matter 
and radiation are concentrated in the big bang. 
It is important to note that $q = -1$ and $\frac{dH}{dt} = 0$ for $t \rightarrow \infty$ in the derived 
model which implies the fastest rate of expansion of universe. So, the derived model can be utilized 
to describe the dynamics of universe at present epoch. Since $lim_{t\rightarrow \infty}\frac{\sigma}{\theta} = 0$, 
thus the model approaches isotropy at late times.\\

\end{document}